\documentclass[twocolumn,showpacs,superscriptaddress,preprintnumbers,amsmath,amssymb,prc]{revtex4-1}
\usepackage{multirow}
\usepackage{amssymb}
\usepackage{amsmath}
\usepackage{graphicx}
\usepackage[normalem]{ulem}
\usepackage{cancel}
\usepackage{multirow}
\usepackage{CJK}
\usepackage[usenames]{color}
\usepackage[colorlinks,linkcolor=blue,urlcolor=blue,citecolor=blue]{hyperref}

\usepackage{tikz,xcolor,hyperref}

\definecolor{lime}{HTML}{A6CE39}
\DeclareRobustCommand{\orcidicon}{
	\begin{tikzpicture}
	\draw[lime, fill=lime] (0,0) 
	circle [radius=0.16] 
	node[white] {{\fontfamily{qag}\selectfont \tiny ID}};
	\draw[white, fill=white] (-0.0625,0.095) 
	circle [radius=0.007];
	\end{tikzpicture}
	\hspace{-2mm}
}
\foreach \x in {A, ..., Z}{%
	\expandafter\xdef\csname orcid\x\endcsname{\noexpand\href{https://orcid.org/\csname orcidauthor\x\endcsname}{\noexpand\orcidicon}}
}

\usepackage{soul,xcolor}
\setstcolor{red}
\newcommand\redsout{\bgroup\markoverwith{\textcolor{red}{\rule[0.5ex]{2pt}{0.4pt}}}\ULon}

\begin{document}
\begin{CJK*} {UTF8} {gbsn}

\title{Probing in-medium effect via giant dipole resonance in the extended quantum molecular dynamics model}

\author{Chen-Zhong Shi(施晨钟)}
\affiliation{Shanghai Institute of Applied Physics,  Chinese Academy of Sciences, Shanghai 201800, China}
\affiliation{National Key Laboratory of Thorium Energy, Shanghai Institute of Applied Physics, Chinese Academy of Science, Shanghai 201800, China}

\author{Xiang-Zhou Cai(蔡翔舟)}
\affiliation{Shanghai Institute of Applied Physics,  Chinese Academy of Sciences, Shanghai 201800, China}%
\affiliation{National Key Laboratory of Thorium Energy, Shanghai Institute of Applied Physics, Chinese Academy of Science, Shanghai 201800, China}

\author{Yu-Gang Ma(马余刚)\orcidC{}} \thanks{Corresponding author:  mayugang@fudan.edu.cn}
\affiliation{Key Laboratory of Nuclear Physics and Ion-Beam Application (MOE), Institute of Modern Physics, Fudan University, Shanghai 200433, China}
\affiliation{Shanghai Research Center for Theoretical Nuclear Physics， NSFC and Fudan University, Shanghai 200438, China} 
\affiliation{School of Physics, East China Normal University, Shanghai 200241, China}

\date{\today}             

\begin{abstract}

Rather than using the geometric method employed in the original Extended Quantum Molecular Dynamics (EQMD) model, this article employs a stochastic approach to analyze the collision term and examine the width of the isovector giant dipole resonance (GDR) in ${}^{208}$Pb.
Based on the ``soft" EQMD model, which we recently developed, the response and strength functions are self-consistently determined for various symmetry energy coefficients and in-medium reduction factor values.
The results confirm that the peak position and GDR width in ${}^{208}$Pb are highly sensitive to the symmetry energy and the in-medium nucleon-nucleon ({\it NN}) cross section. This provides an opportunity to study the nuclear equation of state (EoS) and the medium effect.
A significant reduction in free {\it NN} elastic cross sections within the medium is necessary to accurately reproduce the GDR width, as demonstrated by a comparison with the evaluation data.

\end{abstract}

\maketitle

\section{Introduction\label{introduction}}

With the completion of new-generation nuclear scientific facilities or detection systems worldwide \cite{FRIB,HIAF,Slegs2,Slegs}, a large amount of giant resonance experimental data can be continuously accumulated from heavy-ion \cite{PPNP1}  and photon-induced collisions  \cite{PPNP2,PPNP3,Slegs3,Jiao} in laboratories. A novel way to study the nuclear collective excitations has been proposed, even far, using vortex $\gamma$ rays \cite{Lu,Colo,LiYX}.
Owing to understand the underlying physical mechanism, transport approaches are used to explain the complicated phenomena in HICs.
As an ingredient in the transport approach, the collision term requires the nucleon-nucleon ({\it NN}) cross section as input. 
However, the {\it NN} cross section in free space ($\sigma^\text{free}_{NN}$) is not sufficient for the description of HICs, since the colliding nucleons are surrounded by nuclear matter.
Instead of $\sigma^\text{free}_{NN}$, in-medium cross section $\sigma_{NN}^\ast$, which incorporates medium modifications of the {\it NN} cross section, is usually adopted by transport models.  
Although the cross section in a medium cannot be directly measured experimentally in the same way as in free space, its value can be determined by theoretical studies, such as the Brueckner theories \cite{Brueckner1,Brueckner2} and the closed time path Green's function approach \cite{Green1,Green2}.
Current consensus supports medium effect suppressing the free {\it NN} cross sections, but the magnitude of the reduction is still an open question. 
Experimentally, there are some observables sensitive to the binary collisions that can be used to determine the medium effect on {\it NN} cross sections indirectly, for example, the collective flow observables \cite{Ogilvie,Kumar,Ma_flow1}, nuclear stopping power \cite{LiuJY_stopping,ZhangYX_stopping,ZhangGQ_stopping}, {\it NN} momentum correlation function \cite{momentum_correlation_function,WangTT}, shear viscosity \cite{shear_viscosity,LiuHL,Ma_shear1,Ma_shear2}, and the total reaction cross section \cite{Ma_CRS,Cai1998} etc.
In the earlier investigation, the in-medium cross section was assumed to change linearly with density, but was also found to be relative to isospin-asymmetric and total kinetic energy compared to the free space ones \cite{LiBA_inmed}.
Recently, based on the Lattice BUU (LBUU) model \cite{LBUU}, it was found that the GDR width in $^{208}$Pb is sensitive to the in-medium {\it NN} cross section $\sigma_{NN}^\ast$ since the oscillation damping in resonance is derived mainly from binary collisions for a heavy nucleus system \cite{wangrui2020,SongYD2023,Song2025}.
It helps us to understand how the behavior of the medium affects on the {\it NN} cross section and to verify the unified parameterization of $\sigma_{NN}^\ast$ in the low-energy region.
Of course, it will be certainly interested in the performance of our ``soft" EQMD model which was just recently developed~\cite{shi_gmr,shi_gdr}.

The original EQMD model \cite{EQMD} was presented by Maruyama {\it et al.} in 1996 to study low energy heavy-ion reactions with good computational performance.
In the model, a phenomenological Pauli potential was adopted to mimic Fermion properties. 
In order to obtain a stable ground state of the nucleus, a friction cooling method was employed in the initial stage.
Unlike most QMD-like transport models, the EQMD model adopts dynamical wave packets which are considered including more quantum properties~\cite{FMD}.
In addition, the spurious zero-point center-of-mass kinetic energy has also been subtracted to describe the fragment formation reasonably \cite{AMD1992}.
Based on the above aspects, the EQMD model gives us a special opportunity to research the $\alpha$-cluster effects and the information about nuclear deformation, which are currently very popular in nuclear physics as well as astrophysics \cite{Nat1,Nat2,Nat3,ChenJH}.
Since the model was presented, it has been widely used to investigate HICs at low and intermediate energies, such as collective flow \cite{guocc,shicz2}, giant resonance \cite{HeWB_PRL,HeWB_PRC,WangSS2017,CaoYT2022,EQMD_Bubble,Natowitz}, bremsstrahlung \cite{shicz1,shicz2}, photonuclear reaction \cite{Huang_photonuclear1,Huang_photonuclear2}, liquid-gas phase transition \cite{Wada,guocq,CaoYT2023}, nuclear stopping power \cite{YaoSY}, short range correlation \cite{shenl} and so on.
However, the EQMD model has experienced a long period of development stagnation until recently we updated the mean-field to correct the overestimate on the incompressibility \cite{shi_gmr,shi_gdr}. 
However, for the collision term, it still remains the same as the model was first published in 1996. 

In this paper, we study the GDR oscillation in $^{208}$Pb in the framework of the ``soft" EQMD model with binary collisions via the stochastic method.  
The symmetry energy coefficient and the medium correction in $\sigma_{NN}^\text{free}$ are given by comparison with the evaluation data.
The paper is structured as follows.
A review of the propagation of nucleons in the mean-field potential from the original EQMD model and the ``soft" EQMD model is given in Sect. \ref{EQMD}.
The collision term implemented via the original geometric approach and the stochastic approach is given in Sect. \ref{geometry} and \ref{stochastic}, respectively. 
The GDR oscillation is also described in Sect. \ref{gdr} briefly. 
The results and discussion are presented in Sect. \ref{result}, and a conclusion is given in Sect. \ref{conclusion}.

\section{Model and method \label{method}}

\subsection{Equation of motion\label{EQMD}}

In the framework of the EQMD model, the nucleons are assumed as Gaussian wave packets, and the total wave function is a direct production of nucleons as follows:
\begin{equation}\label{wave}
\begin{aligned}
\Psi & =  \prod_i \varphi_i \left(\mathbf{r}_i\right) \\ & =\prod_i \left(\frac{v_i+v_i^*}{2 \pi}\right)^{3 / 4} \exp \left[-\frac{v_i}{2}\left(\mathbf{r}_i-\mathbf{R}_i\right)^2+\frac{i}{\hbar} \mathbf{P}_i \cdot \mathbf{r}_i\right].
\end{aligned}
\end{equation}
Here $\mathbf{R}_i$ and $\mathbf{P}_i$ are the position and moment of the center of wave packet belonging to the $i$-th nucleon in phase spaces. 
$v_i=1/\lambda_i+i\delta_i$ is the complex width, where $\lambda_i$ and $\delta_i$ are the real and imaginary parts, respectively.
Its corresponding density distribution can be written as 
\begin{equation}\label{eq:density}
\rho_i \left( \mathbf{r} \right) = \frac{1}{\left( \pi \lambda_i \right)^{3/2}} \exp \left[ -\left( \mathbf{r}-\mathbf{R}_i\right)^2/ \lambda_i \right].
\end{equation}
Under the time dependent variation principle (TDVP) \cite{FMD}, the propagation of particles can be described as 8-$A$ dimensional classical equations of motion as follows:
\begin{equation}
\begin{aligned}
\dot{\mathbf{R}}_i & =\frac{\partial H}{\partial \mathbf{P}_i}+\mu_{\mathrm{R}} \frac{\partial H}{\partial \mathbf{R}_i}, ~~\dot{\mathbf{P}}_i=-\frac{\partial H}{\partial \mathbf{R}_i}+\mu_{\mathrm{P}} \frac{\partial H}{\partial \mathbf{P}_i}, \\
\frac{3 \hbar}{4} \dot{\lambda}_i & =-\frac{\partial H}{\partial \delta_i}+\mu_\lambda \frac{\partial H}{\partial \lambda_i}, ~~\frac{3 \hbar}{4} \dot{\delta}_i=\frac{\partial H}{\partial \lambda_i}+\mu_\delta \frac{\partial H}{\partial \delta_i},
\end{aligned}
\end{equation}
where $\mu_\mathbf{R}$, $\mu_\mathbf{P}$, $\mu_{\lambda}$ and $\mu_{\delta}$ are the corresponding friction coefficients. 
In the initial state, they are negative values that dissipate the energy of the total system down to its minimum point as the ground state~\cite{EQMD}.
On the other side, they remain strictly zero to satisfy the conservation of total energy.
It is worth noting that the width of each wave packet is treated as a dynamical variable, which is considered as a semi-classical, semi-quantum mechanics involves more quantum effects than the fixed-width situation~\cite{FMD}.
$H$ represents the Hamiltonian expressed as
\begin{equation}\label{eq:hamiltonian}
\begin{aligned}
H & =\left\langle\Psi\left|\sum_i-\frac{\hbar^2}{2 m} \nabla_i^2-\hat{T}_{\text {zero }}+\hat{H}_{\text {int }}\right| \Psi\right\rangle \\
& =\sum_i\left[\frac{\mathbf{P}_i^2}{2 m}+\frac{3 \hbar^2\left(1+\lambda_i^2 \delta_i^2\right)}{4 m \lambda_i}-\frac{t_i}{M_i}\right]+H_{\mathrm{int}} .
\end{aligned}
\end{equation}
Here, the sum of the three terms in square brackets is the total kinetic energy belonging to the $i$-th particle.
The first term represents the kinetic energy of the center of the wave packet.
The second term arises from the momentum distribution. 
And $t_i$ is the zero-point kinetic energy, which is defined as follows:
\begin{equation}
t_i
 = \frac{\langle\phi_i|\hat{\mathbf{p}^2}| \phi_i\rangle}{2 m}-\frac{\langle\phi_i|\hat{\mathbf{p}}| \phi_i\rangle^2}{2 m}.
\end{equation}
$M_i$ is the namely ``mass number" of the fragment to which the $i$-th nucleon belongs. 
It can be written as 
\begin{equation}\label{eq:mass number}
\begin{aligned}
M_i  &=\sum_j F_{i j}, \\
F_{i j} & =\left\{\begin{array}{cc}
1, & \left|\mathbf{R}_i-\mathbf{R}_j\right|<a, \\
e^{-\left(\left|\mathbf{R}_i-\mathbf{R}_j\right|-a\right)^2 / b}, & \left|\mathbf{R}_i-\mathbf{R}_j\right| \geqslant a.
\end{array}\right.
\end{aligned}
\end{equation}
$H_\text{int}$ represents the total potential energy consisting of effective interaction, Coulomb interaction, and Pauli potential, i.e., 
\begin{equation}\label{eq:mean field potential}
H_\text{int.}=\int{u(\rho)\text{d}\mathbf{r}}+H_\text{Coul.}+H_\text{Pauli}.
\end{equation} 
In the most original EQMD model, a very simple form of effective interaction was adopted, leading to very hard incompressibilities that confuse with the current consensus. 
Recently, we overcome this confusion by employing a typical template of the Skyrme energy density functional with ``soft" interaction parameters \cite{shi_gmr,shi_gdr,Feng}.
The energy density functional can be expressed as follows:
\begin{equation}\label{eq:new}
\begin{aligned}
u(\rho)&=  \frac{\alpha}{2} \frac{\rho^2}{\rho_0}+\frac{\beta}{(\gamma+1) \rho_0^\gamma} \rho^{\gamma+1}\\
&+\frac{g_{\text {Surf.}}}{2 \rho_0}(\nabla \rho)^2+\frac{g_{\text {Surf.}}^{\text {iso }}}{2 \rho_0}\left[\nabla\left(\rho_{\mathrm{n}}-\rho_{\mathrm{p}}\right)\right]^2 \\
& +\frac{c_s}{2 \rho_0^{\gamma_s}} \rho^{\gamma_s-1} \left(\rho_\text{n}-\rho_\text{p}\right)^2+g_\tau \frac{\rho^{8 / 3}}{\rho_0^{5 / 3}}.
\end{aligned}
\end{equation} 
The coefficients $\alpha$, $\beta$, and $\gamma$ are related to the bulk term.
$g_\text{Surf.}$ and $g_\text{Surf.}^\text{iso}$ are the coefficients related to the surface interaction and its symmetry part, respectively.
$c_\text{s}$ is the symmetry energy coefficient.
$\gamma_s$ is taken as 0.5 \cite{gamma_05,double_ratio_np_buu}, which represents a soft-asy EoS.
The last term in eq.~\ref{eq:new} corresponds to the momentum-dependent interaction and $g_{\tau}$ is the corresponding coefficient. 
Here, we adopt a widely used formation of the momentum-dependent interaction \cite{Feng_2011,Zhang_2012}, derived from the standard Skyrme interaction via the local density approximation for uniform nuclear matter at zero temperature.
The Coulomb interaction utilized in this article consists of direct and exchange parts as follows:
\begin{equation}
H_{\text{Coul.}}(r) = \frac{a}{2} \int \mathrm{d}^3 r^{\prime} \frac{\rho_{\mathrm{p}}(r)\rho_{\mathrm{p}}\left(r^{\prime}\right)}{\left|r-r^{\prime}\right|}-\frac{3}{4} e^2\left(\frac{3}{\pi}\right)^{1 / 3}  \rho_{\mathrm{p}}^{4 / 3}(r),
\end{equation}
where $a$ is the fine structure constant.
The Pauli potential $H_\text{Pauli}$ is defined as
\begin{equation} \label{eq:pauli}
H_\text{Pauli}  = \frac{c_\text{P}}{2} \sum_i\left(f_i-f_0\right)^\mu \theta\left(f_i-f_0\right),
\end{equation}
where $c_\text{P}$ and $\mu$ are the strength and power parameters, respectively.
In Eq.~\ref{eq:pauli}, it takes 12.0 MeV for $c_\text{P}$ and 1.3 for $\mu$.
$\theta$ is the unit step function and $f_0$ is the corresponding threshold parameter, which usually takes a value close to 1. 
The $f_i$ represents the overlap of the $i$-th nucleon with all identical particles defined as follows:
\begin{equation} \label{eq:overlap}
f_i  \equiv \sum_j \delta\left(S_i, S_j\right) \delta\left(T_i, T_j\right)\left|\left\langle\phi_i \mid \phi_j\right\rangle\right|^2.
\end{equation}
The Pauli potential mimics the fermion properties of nuclear matter in a low-energy region, which can be regarded as a repulsive force that prohibits the identical particles from being closed to each other in the phase space.

In this work, we use the SLy7 set of parameters for the calculation, and the details are given in the tab.~\ref{tab:table_SLy7}.

\begin{table}[t]
\caption{\label{tab:table_SLy7}}
Parameters of SLy7 and some corresponding physical variables.
\begin{ruledtabular}
\begin{tabular}{lr}
$\alpha$ (MeV)               & -294.0\\ 
$\beta$ (MeV)                & 215.0 \\
$\gamma$                     & 7/6   \\
$g_\text{Sur.}$ (MeV)        & 22.6  \\
$g_\text{Sur.}^{iso}$ (MeV)  & -2.3  \\
$g_{\tau}$ (MeV)             & 9.9   \\
$c_\text{s}$ (MeV)           & 32.6  \\
$\rho_{\infty}$ (fm$^{-3}$)  & 0.158 \\
$m^{\prime}/m$               & 0.69  \\
$K_\infty$ (MeV)             & 229   \\
\end{tabular}
\end{ruledtabular}
\end{table}

\subsection{Two-body collision \label{geometry}}

Binary collision is another important aspect for transport models.
During every time step $\delta t$, {\it NN} scattering is tested for each pair of particles.
Before we introduce our method, a brief review of the collision term in the original EQMD model is given here.

Similarly to Bertsch's prescription \cite{Bertsch}, there are two necessary conditions that must be satisfied, then the scattering can happen.
The first is that the relative distance between the colliding nucleons 1 and 2 must take the minimum value within a time step $[t,t+\delta t]$.
Generally speaking, $0.5~\text{fm}/c$ is a typical value of $\delta t$ for the low-intermediate energies reaction.   
The second is that the minimum distance must be smaller than a threshold value, which takes $d_\text{coll}=2.0~\text{fm}$ in the EQMD model.
If such two conditions are satisfied, a collision is attempted with probability $P_\text{coll}$ as
\begin{equation} \label{eq:collision probability}
P_\text{coll}=\frac{\sigma_{NN}}{\sigma_\text{max}}.
\end{equation}
Here, $\sigma_\text{max}=\pi d^2_\text{coll}$.
$\sigma_{NN}$ is taken an energy-dependent cross section \cite{EQMD, AMD1992}, which is represented by $\sigma_{NN}^\text{EQMD}$ in the text as
\begin{equation} \label{eq:original EQMD cross section}
\begin{aligned}
\sigma_{NN}^\text{EQMD}=&\frac{100~\text{mb}}{1+\epsilon/200~\text{MeV}},\\
\epsilon \equiv& \frac{\textbf{p}^2_\text{rel}}{2m}.
\end{aligned}
\end{equation}
Here $\textbf{p}_\text{rel}=\textbf{P}_1-\textbf{P}_2$ is the relative momentum of the center of wave packets 1 and 2.
In the early AMD model \cite{AMD1992}, such a cross-section design was used to ensure that the collision probability was less than 1.
Subsequent AMD updates \cite{AMD1993,AMD_PPNP} have used a more complex parameterization.
In the EQMD model, an isotropic angular distribution is assumed for the outgoing nucleons, and their relative momentum is adjusted iteratively until the final system satisfies the conservation of energy due to the Pauli potential, which also depends on the nucleon momenta.
If such a value for the relative momentum does not exist in the final system, the scattering is regarded as blocked.
Besides, there is an additional restriction for the binary collision in the framework of the EQMD model.
It requires that the collision energy of the two nucleons should be greater than 1 MeV in their center-of-mass system.
Although it is reported that its effect on the final results is usually expected to be unimportant \cite{AMD_PPNP}.

After an attempted collision, the fermion properties are also checked.
The so-called occupation probability of the outgoing nucleons, which is defined as 
\begin{equation} \label{eq:occupation}
(P_\text{occ.})_i=\text{min}\left( 1, f_i^\prime-f_0\right).
\end{equation}
Here, $i=1,~2$ represents the scattered nucleons 1 and 2.
$f_i^\prime$ are their overlaps calculated with new centroid momenta $\textbf{P}_i^\prime$ according to eq.~\ref{eq:overlap}.
And $f_0$ is the same parameter as in Eq.~\ref{eq:pauli}.
Then the probability that Pauli-blocking occurred can be obtained as follows:
\begin{equation} \label{eq:Pauli blocking probability}
P_\text{blocking}=1-[1-(P_\text{occ.})_1][1-(P_\text{occ.})_2].
\end{equation}
If a blocking occurred, the outgoing nucleons would restore their initial states.

Clearly, the two-body collision term used in the original EQMD is a  typical geometric approach.
This approach is widely used in transport models for binary collisions.
However, recent studies \cite{Xu_PPNP} have reported some shortcomings in this method.
For example, there is a 50\% chance of spurious collisions arising for the same pair of collided particles in the subsequent time step.
One possible solution is to prevent the same pair from colliding again unless they have scattered with other particles.
Another issue is that the $\sigma_\text{max}$ in Eq.~\ref{eq:collision probability} must be larger than the {\it NN}  cross section to ensure the collision probability is less than 1; this is an arbitrary constant.
Similarly, the minimum distance $d_\text{coll}$ is subject to the same issue.

\subsection{Stochastic method \label{stochastic}}

In view of the complexity in the detail of the geometric approach utilized by different groups \cite{box}, we introduce a more reliable treatment, the so-called stochastic method, which can be considered as the mean free path method \cite{Xu_PPNP}, into our ``soft" EQMD model.
For colliding particles 1 and 2, we assume that the probability of scattering is proportional to their density overlap, as described in \cite{AMD_PPNP}:
\begin{equation}\label{eq:density overlap}
\int \rho_1 \rho_2 \mathrm{d} \mathbf{r}=\left[\frac{1}{\left(\lambda_1+\lambda_2\right) \pi}\right]^{\frac{3}{2}} \exp\left[ -\frac{ \left(\mathbf{R}_1-\mathbf{R}_2\right)^2}{\lambda_1+\lambda_2}\right],
\end{equation}
and the collision vertex $\mathbf{r}_{1,2}$ can be sampled according to the following distribution \cite{shicz2}:
\begin{equation}\label{eq:vertex}
\mathbf{r}_{1,2}  \sim \frac{\rho_1 \rho_2}{\int \rho_1 \rho_2 d \mathbf{r}}. 
\end{equation}
If the colliding nucleons move in a straight line within the time interval $\left[t_0, t_0+\delta t\right]$, the probability of scattering occurring can be expressed as follows:
\begin{equation}\label{eq:stochastic probability}
\begin{aligned}
P_\text{coll}=&\frac{\sigma_{NN}}{\left(\lambda_1+\lambda_2\right) \pi} \exp \left[-\frac{d_\perp^2}{\lambda_1+\lambda_2}\right] \\
&\times  \sqrt{\frac{1}{\left(\lambda_1+\lambda_2\right) \pi}} \int_{t_{0}}^{t_0+\delta t} \exp \left[-\frac{d_\parallel^2}{\lambda_1+\lambda_2}\right] v_\text{rel} \mathrm{d}t.
\end{aligned}
\end{equation}
Here, $d_\parallel$ and $d_\perp$ are the distances parallel and perpendicular to the relative velocity $v_\text{rel}$ of the centers of their wave packets, respectively.
$\sigma_{NN}$ is taken as the free {\it NN} elastic scattering cross section obtained from Ref.~\cite{Cugnon}, which is represented by $\sigma^\text{free}_{NN}$ in the text.
For the pp/nn system, the elastic cross section can be written as
\begin{equation}\label{eq:cugnon_pp}
\begin{aligned}
\sigma_{\mathrm{pp/nn}} &=628.367, & & p_\text{lab}<0.1,\\
&= 34\left( \frac{p_\text{lab}}{0.4} \right)^{-2.104}, & & 0.1<p_\text{lab}<0.4,\\ 
& =23.5+1000\left(p_{\mathrm{lab}}-0.7\right)^4, & & 0.4<p_{\mathrm{lab}}<0.8, \\
& =\frac{1250}{p_{\mathrm{lab}}+50}-4\left(p_{\mathrm{lab}}-1.3\right)^2, & & 0.8<p_{\mathrm{lab}}<2.0, \\
& =\frac{77}{p_{\mathrm{lab}}+1.5}, & & 2.0<p_{\mathrm{lab}}.
\end{aligned}
\end{equation}
For the pn system, the elastic cross section can be written as
\begin{equation}\label{eq:cugnon_pn}
\begin{aligned}
\sigma_{\mathrm{pn}} =&3627.89, ~~~~~~~~~~~~~~~~~~~~~~~~~~~~~~~~~~~ p_\text{lab}<0.05,\\
=&  6.3555 p_\text{lab}^{-3.2481} \mathrm{exp} \left( -0.377 \left( \text{In}\left(p_\text{lab}\right)\right)^2 \right), \\ 
&~~~~~~~~~~~~~~~~~~~~~~~~~~~~~~~~~~~~~~~~0.05<p_\text{lab}<0.4,\\
=&33+196\left|p_{\mathrm{lab}}-0.95\right|^{2.5},~~~~~~~~~0.4<p_{\mathrm{lab}}<0.8, \\
=&\frac{31}{\sqrt{p_{\mathrm{lab}}}},~~~~~~~~~~~~~~~~~~~~~~~~~~~~~~~~0.8<p_{\mathrm{lab}}<2, \\
=&\frac{77}{p_{\mathrm{lab}}+1.5},~~~~~~~~~~~~~~~~~~~~~~~~~~~~~2<p_{\mathrm{lab}}.
\end{aligned}
\end{equation}
Since there is scarce experimental data for the lower incident energy, we fix the value of $\sigma_\text{pp} \left( 0.1~\text{GeV}/c\right)$ for pp/nn system at $p_\text{lab}<0.1~\text{GeV}/c$ and $\sigma_\text{pn} \left(0.05~\text{GeV}/c\right)$ for the pn system at $p_\text{lab}<0.05~\text{GeV}/c$ same as Ref.~\cite{wangrui2020}.

\begin{figure}[htbp]
\resizebox{8.6cm}{!}{\includegraphics{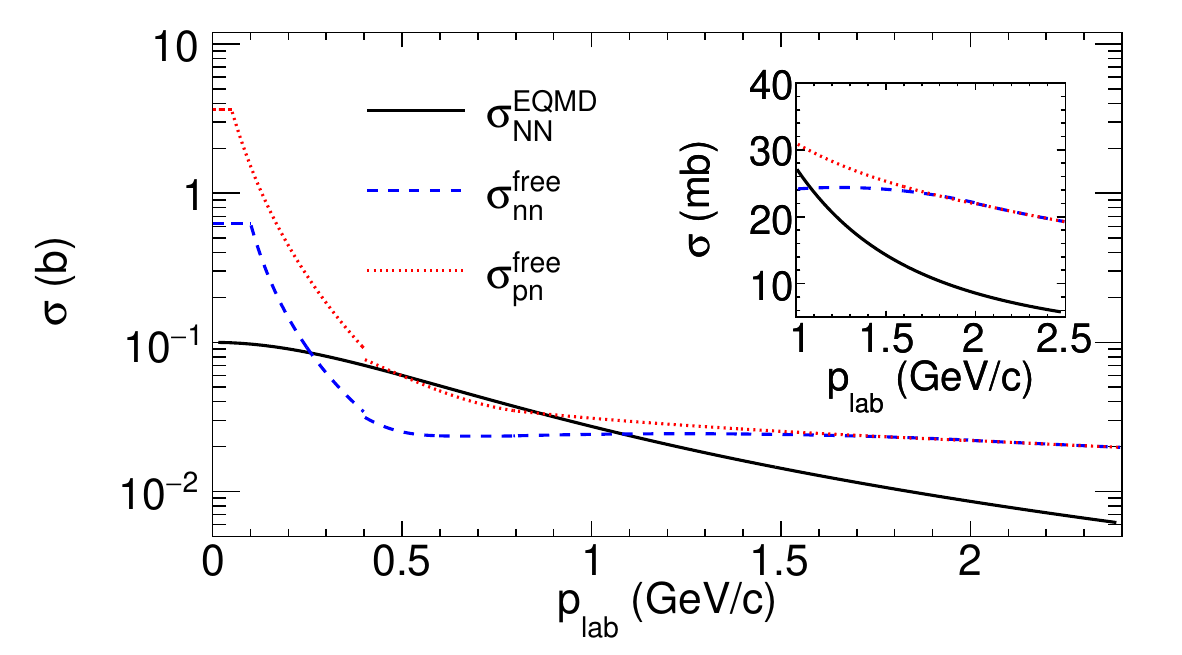}}
\caption{\label{fig:cugnon cross section} Comparison of the {\it NN} elastic cross sections utilized in the original EQMD and the parameterization of elastic cross-sections given by Ref.~\cite{Cugnon}.
Note the different scales in the inset.}
\end{figure}

In Fig.~\ref{fig:cugnon cross section}, the free $\it{NN}$ elastic cross-sections for $p_\text{lab} \leq 2.5~\text{GeV}/c$ is plotted in logarithmic coordinates. 
The inset shows the same thing for $1.0 \leq p_\text{lab} \leq 2.5~\text{GeV}/c$ in linear coordinates.
The dashed line and the dot line represent the parameterization of Eq.~\ref{eq:cugnon_pp} and Eq.~\ref{eq:cugnon_pn} for the pp/nn and pn systems obtained from Ref.~\cite{Cugnon}.
The solid line is the elastic cross section employed in the original EQMD for comparison.
When the incident momentum $p_\text{lab}$ is smaller than about $0.1~\text{GeV}/c$, i.e., 2.5 MeV $\it{NN}$ cm energy, $\sigma_\text{pn}$ is significantly larger than in the other situations.
According to Eq.~\ref{eq:original EQMD cross section}, 100 mb is the maximum cross section in the original EQMD, which is about 3 b lower than $\sigma_\text{pn}$ and about 500 mb lower than $\sigma_\text{pp}$ in the low energy region. 
However, as the incident momentum increases, the differences narrowed rapidly.
The inset shows that the difference in cross section reduces to about several tens of mb when $p_\text{lab}$ is greater than about $1~\text{GeV}/c$. 

Note that the calculation of relative momentum between particles 1 and 2 in the scattering method differs from the most QMD-like approach.  
In our previous works \cite{shicz1,shicz2}, it was found that the momenta of colliding particles should be extra considered since the specific definition of the total kinetic energy in the Hamiltonian, i.e., Eq.~\ref{eq:hamiltonian}.
In order to successfully extract the incident energy, we adopted the method suggested in our previous article.
The Wigner function of the particle $i$ can be written as 
\begin{equation}
\begin{aligned}
w_i(\mathbf{r}, \mathbf{p})= & \left(\frac{1}{\pi \hbar}\right)^3 \exp \left[-\frac{1+\lambda_i^2 \delta_i^2}{\lambda_i}\left(\mathbf{r}-\mathbf{R}_i\right)^2\right. \\
& \left.-\frac{2 \lambda_i \delta_i}{\hbar}\left(\mathbf{r}-\mathbf{R}_i\right)\left(\mathbf{p}-\mathbf{P}_i\right)-\frac{\lambda_i}{\hbar^2}\left(\mathbf{p}-\mathbf{P}_i\right)^2\right] .
\end{aligned}
\end{equation}
In many ways, its behavior is the closest analogue to a classical phase-space density.
Equation~\ref{eq:vertex} shows the collision vertex of particles 1 and 2.
When combined with their Wigner functions and the kinetic energy formula, their momenta can be obtained according to the following distribution:
\begin{equation}
\begin{aligned}
\mathbf{p}_i &\sim \left(\frac{\lambda_i}{\pi \hbar^2}\right)^{\frac{3}{2}} \exp \left\{-\frac{\lambda_i}{\hbar^2}\left[\mathbf{p}_i-\left(\mathbf{P}_i-\hbar\delta_i \mathbf{r}_{1,2}+\hbar\delta_i \mathbf{R}_i\right)\right]\right\},\\
\mathbf{p}_i^\prime&=\mathbf{P}_i+\left(\mathbf{p}_i-\mathbf{P}_i\right)\sqrt{1-\frac{1}{M_i}}
\end{aligned}
\end{equation}
where $i$ = 1, 2 represent particles 1 and 2.
$M_i$ is the same as in Eq.~\ref{eq:mass number}.
$\mathbf{p}_i^\prime$ is the momentum re-extracted that belongs to the $i$-th particle.
This treatment significantly impacts the calculation of bremsstrahlung.
For the EQMD model, the experimental data can only be reasonably reproduced by considering such a treatment.
Further details can be found elsewhere (see references~\cite{shicz1,shicz2}).

At the end of the subsection, we give the proof that our method is equivalent to the mean free path method when the real part of the wave packet width belonging to the incident particle approaches $0$.
If we assume nucleon 1 as the incident particle, its density distribution function (see Eq.~\ref{eq:density}) becomes a delta function, i.e., $\rho_1\left(\mathbf{r}\right) \rightarrow \delta(\mathbf{r}-\mathbf{R}_1)$, when $\lambda_1 \rightarrow 0$.
The value of the density overlap of nucleons 1 and 2 is $\rho_2\left( \mathbf{R}_1 \right)$ obtained by Eq.~\ref{eq:density overlap}.
According to the mean value theorem of integrals, the density overlap integrated over $d_\parallel$ in the time step $\left[t_0,t_0+\delta t \right]$ can be written as 
\begin{equation}\label{eq:stochastic 2}
P_\text{coll}=\sigma_\text{NN} \bar{\rho} \mathbf{v}_\text{rel} \delta t,
\end{equation}
Here, $\bar{\rho}$ is the average linear density on the trajectory of the incident particle.
From Ref.~\cite{Xu_PPNP}, the mean free path is known to be expressed as $1/\left(\sigma \rho\right)$, where $\rho$ is the number density of the incident particle experienced in a small volume.
The relaxation time $\tau$ for binary collision is $\lambda/\mathbf{v}_\text{rel}$.
The probability of scattering in the time interval $\delta t$ is therefore $\delta t/\tau=\left(\mathbf{v}_\text{rel}/\lambda\right)\delta t=\sigma \rho \mathbf{v}_\text{rel} \delta t$.
The probability of two-body scattering is thus equivalent to Eq.~\ref{eq:stochastic 2} when the width of the incident particle approaches 0. 

Besides, in actual calculations, to prevent the possibility of a collision probability greater than 1, we adopt the following treatment
\begin{equation}
P^\prime_\text{coll} = 1-e^{-P_\text{coll}}.
\end{equation}
Here, $P_\text{coll}$ is obtained from Eq.~\ref{eq:stochastic probability} and $P^\prime_\text{coll}$ is used to determine whether a collision has occurred.
A similar treatment has been employed by the Constrained Molecular Dynamics (CoMD) model~\cite{box}.
To our knowledge, the only QMD-like model that utilizes the stochastic method is the CoMD model.

\subsection{ Isovector giant dipole resonance \label{gdr}}

The isovector giant dipole resonance can be regarded as a collective oscillation between centroids of protons and neutrons with opposite phases. 
In order to excite a nucleus undergoing such a collective oscillation, a perturbation operator can be added to the Hamiltonian at $t_0$, which can be written as
\begin{equation}
\label{eq:Q_gdr}
\begin{aligned}
\hat{Q}_\text{exc.}=&\lambda\hat{Q}\delta\left(t-t_0\right),\\
\hat{Q}=\sum^{A}_{i}{\hat{q}}=&\frac{N}{A}\sum_{i\in\text{p}}{\hat{z}_i}-\frac{Z}{A}\sum_{i\in\text{n}}{\hat{z}_i}.
\end{aligned}
\end{equation}
Here $\hat{Q}$ represents the dipole excitation operator and $\lambda$ is a tiny variable, here it takes 25 MeV/$c$ following Ref.~\cite{Urban}, to ensure the perturbation condition.
Following linear response theory, the response of the excited nucleus can be expressed as a time-dependent function as
\begin{equation}
\label{eq:std}
\begin{aligned}
\Delta\langle\hat{Q}\rangle(t) & = \langle\hat{Q}\rangle\left(t\right)- \left\langle 0|\hat{Q}| 0\right\rangle \\
& =-\frac{2 \lambda \theta(t)}{\hbar} \sum_f|\langle f|\hat{Q}| 0\rangle|^2 \sin \frac{\left(E_f-E_0\right) t}{\hbar}.
\end{aligned}
\end{equation}
Here $|0>$, $|f>$ are the nuclear state before and after perturbation, and $E_0$, $E_f$ are the corresponding eig-energy.
Since $\langle\hat{Q}\rangle$ only explicitly contains the z component, the new momentum of the wave packets is changed after perturbation \cite{Urban} as 
\begin{equation}
(P_{i})_{z} \rightarrow \begin{cases}
(P_i)_z-\lambda \frac{N}{A}, & i \in \text{proton}\\
(P_i)_z+\lambda \frac{Z}{A}, & i \in \text{neutron}
\end{cases}
\end{equation}
The GDR width is defined by the full width at half maximum (FWHM) of the strength function, which can be obtained via a Fourier integral of $\Delta\langle\hat{Q}\rangle(t)$ as follows:
\begin{equation}
\label{eq:se}
S(E_\gamma) = -\frac{1}{\pi \lambda} \int_0^{\infty} \text{d} t \Delta\langle\hat{Q}\rangle(t) \sin \frac{E_\gamma t}{\hbar}.
\end{equation}
Here $E_\gamma$ is the photon emission energy. 

In reality, an artificial damping factor of $e^{-\gamma t/\hbar}$ is multiplied by $\Delta\langle\hat{Q}\rangle(t)$ as usual for the calculation of the strength function.
This is a common practice for studying nuclear giant resonance using transport approaches, e.g., EQMD \cite{HeWB_PRL,WangSS2017}, IBUU \cite{Kong_gdr}, and LBUU \cite{LBUU}.
A purpose of this treatment is to widen the GDR width to avoid spurious oscillations due to the integration of finite-time span in Eq.~\ref{eq:se}.
However, the collisional dissipation hidden in the GDR width has been ignored. 

\section{Results and Discussion \label{result}}

\begin{figure}[htbp]
\resizebox{8.6cm}{!}{\includegraphics{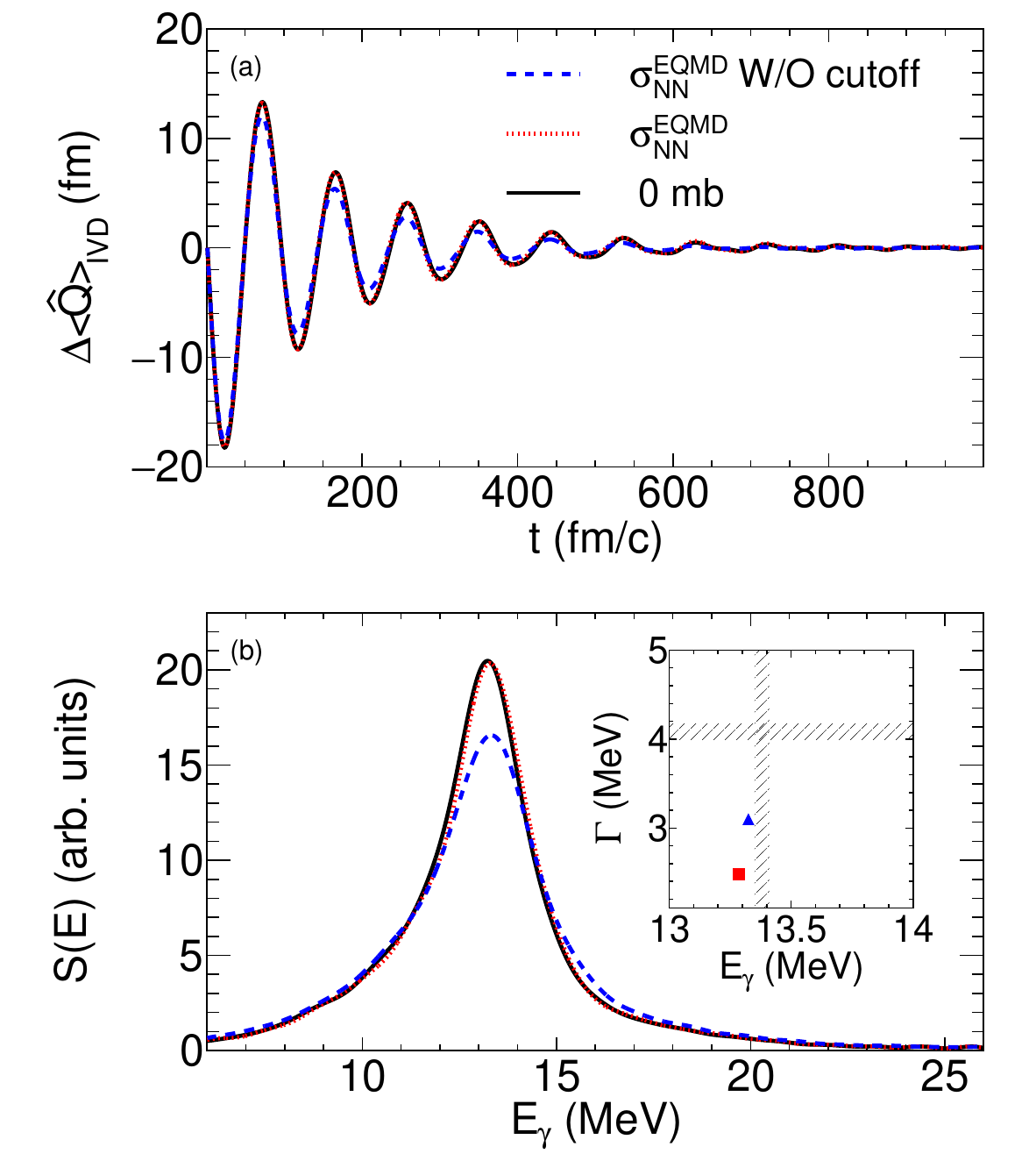}}
\caption{\label{fig:geometry}
The GDR oscillation of $^{208}$Pb as a function of time (a) and the corresponding strength function (b). The peak positions and GDR widths calculated with and without low-energy cut are plotted as a square and a triangle in the inset.
And the shaded area is the evaluation data of GDR peak position and width.}
\end{figure}

To avoid introducing complications that would interfere with the research objective, our previous studies only considered the mean-field aspect~\cite{shi_gmr,shi_gdr}.
This article also takes the collision term into account, enabling the GDR width of $^{208}$Pb to be obtained naturally.

First, we show the GDR oscillation in the framework of our ``soft" EQMD model with the original geometric approach for the {\it NN} scattering.
In Fig.~\ref{fig:geometry}, the response of $^{208}$Pb as a function of time is plotted in the upper panel (a) and the corresponding strength function is drawn in the lower panel (b).
As a benchmark, the solid line is calculated without any binary collision during the GDR oscillation, i.e. $\sigma_{NN}=0~\text{mb}$ is considered in this situation.
For the dotted line and dashed line, we employ the geometric approach with an energy-dependent cross section $\sigma_{NN}^\text{EQMD}$, with and without the low-energy truncation introduced in subsection \ref{geometry}, respectively.
To illustrate more clearly, the corresponding peak positions and GDR widths are plotted as a square and a triangle in the inset.
The hatched bands are the evaluation data calculated by the SLO model from Ref.~\cite{evaluation_data}. 
For peak positions, all situations are very close to evaluation data and only slightly lower than evaluation values.
Compared with the situation with no binary collisions, the difference in the original geometric approach used in the EQMD model is not significant. 
However, when the low-energy cutoff condition is removed from the $NN$ collision, a small damping in the response function can clearly be observed. 
In terms of GDR width, the calculation without the low-energy cutoff is also wider. This suggests that a considerable number of collisions should occur in the low-energy region for dipole oscillations. 
However, using the geometric method employed in the original EQMD model, it is impossible to reproduce the experimental result, regardless of whether the low-energy cutoff is considered.
Our calculations underestimate the GDR width by at least 1 MeV.
However, these results do not negate the effectiveness of the geometric approach to treating binary collisions, as adopted by most BUU- and QMD-like models. They do, however, suggest that the collision term used in the original EQMD is not suitable for reproducing the GDR width in the low-energy region, given that it has not been updated for some time.

\begin{figure}[htbp]
\resizebox{8.6cm}{!}{\includegraphics{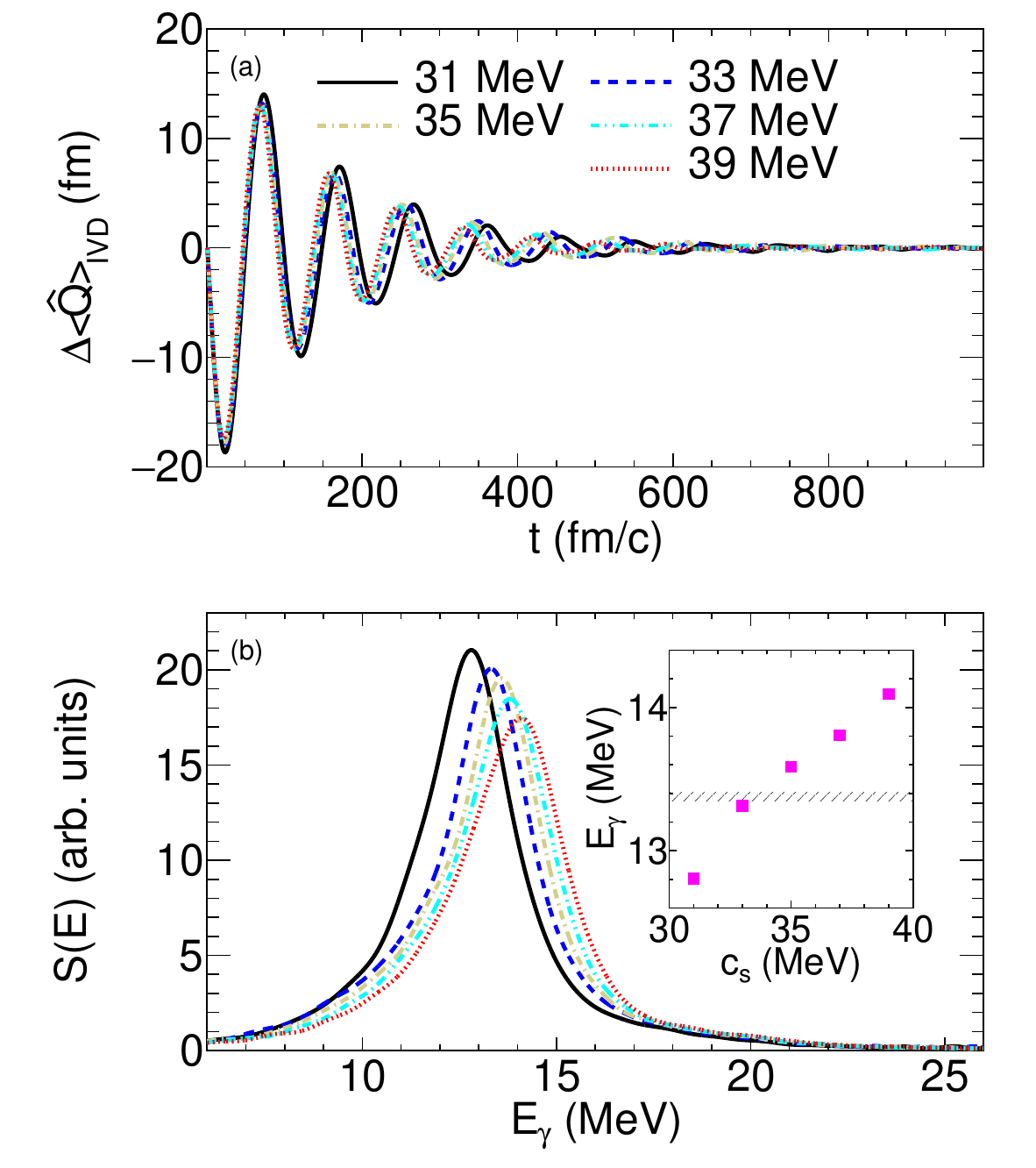}}
\caption{\label{fig:symmetry energy vs peak energy} 
The time evolution of the GDR oscillation for ${}^{208}$Pb with different symmetry energy coefficients (a) and the symmetry energy coefficient dependence of the GDR peak position (b). The corresponding peak positions are plotted as squares in the insert and the shaded area represents the evaluated data. } 
\end{figure}

Fig.~\ref{fig:symmetry energy vs peak energy} shows our study of the dependence of  the GDR peak on the symmetry energy coefficient. 
As can be seen in the upper panel (a), a larger $c_\text{s}$ results in a faster GDR oscillation frequency.
The corresponding strength functions are shown in the lower panel (b).
As an observable sensitive to the symmetry energy, which acts as a restoring force for the GDR, the peak positions plotted as squares in the inset demonstrate an upward trend as $c_\text{s}$. 
Compared with the evaluated data, it appears that $c_\text{s}$ must have a value close to 33 MeV. 
Please note that no collision term has been considered in Fig.~\ref{fig:symmetry energy vs peak energy} to eliminate any influence on the peak position.
The subsequent text shows that the binary collisions affecting the peak position are limited.

\begin{figure}[t]
\resizebox{8.6cm}{!}{\includegraphics{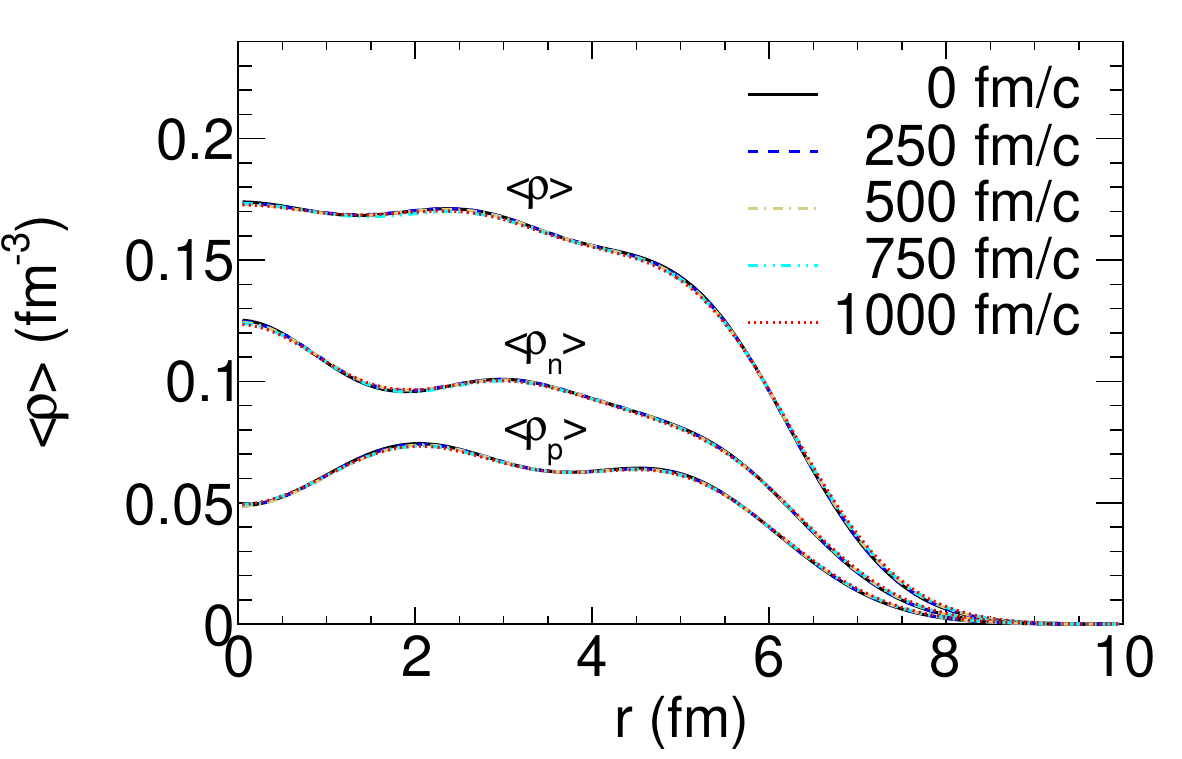}}
\caption{The time evolution of the radial distribution of the density of the ground state $^{208}$Pb nucleus with binary collisions during 1000 fm/$c$.\label{fig:density distribution}}
\end{figure}

To evaluate the influence of the stochastic method on the stability of the ground state, the time evolution of the radial density distribution considered with {\it NN} scattering is plotted in Fig.~\ref{fig:density distribution}.
The nucleus's evolution continues to 1000 fm/$c$ and the corresponding density radius distributions are plotted at 250 fm/$c$ intervals.
As discussed in Sect.~\ref{EQMD}, the ground state is obtained self-consistently using the friction cooling method and a set of SLy7 parameters within the framework of the ``soft" EQMD model.
As shown in the figure, only tiny fluctuations can be observed between each time interval.
The radial density distributions of the nucleus $\langle \rho \rangle$, proton $\langle \rho_\text{p} \rangle$, and neutron $\langle\rho_\text{n}\rangle$ remain stable even if the nucleus undergoes a thousand fm/$c$ evolution.
This demonstrates that the stochastic method employed in our model does not have a negative impact on the stability of the nucleus, making it a valuable tool for the study of long-term nuclear reactions.

\begin{figure}[t]
\resizebox{8.6cm}{!}{\includegraphics{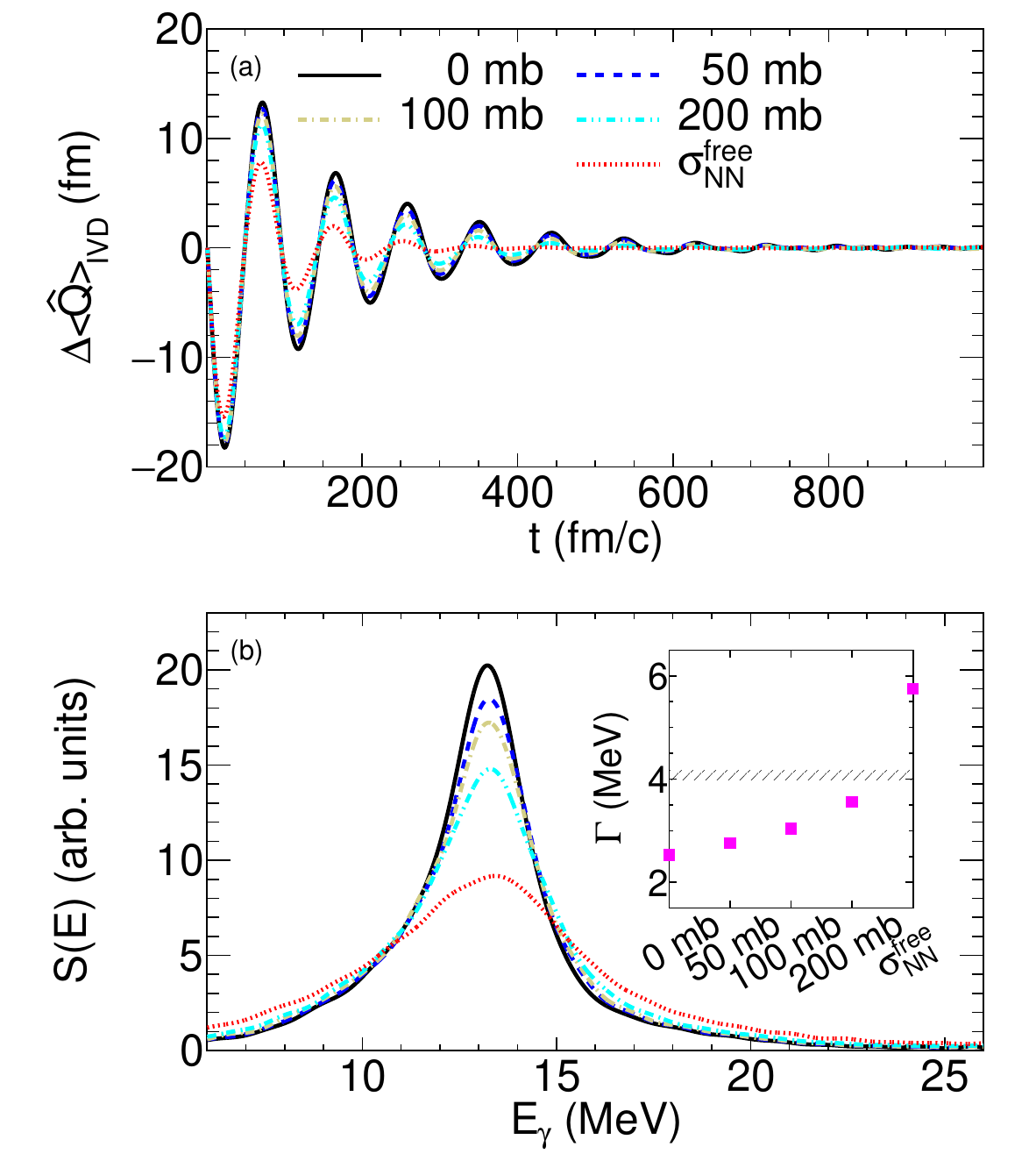}}
\caption{\label{fig:different cross section} The responses of GDR oscillations with constant nucleon-nucleon cross section and the more realistic energy-dependent nucleon-nucleon cross section in panel (a). The corresponding strength functions in panel (b). The inset displays the corresponding decay width and the evaluated data.} 
\end{figure}

As shown in Fig.~\ref{fig:geometry}, the geometric approach used in the original EQMD cannot produce reasonable collisional damping of the GDR oscillation.
However, the situation is very different when the stochastic approach is implemented.
As can be seen in the upper panel (a) of Fig.~\ref{fig:different cross section}, the damping of the dipole oscillation is significantly dependent on the NN cross section.
For comparison, the simulation without any binary collision, i.e. $\sigma_{NN}=0~\text{mb}$, is also displayed.
To illustrate the effect of binary collisions on the GDR oscillation more clearly, the corresponding strength functions are shown in the lower panel (b).
It is evident that an increase in the GDR width depends on an increase in the {\it NN}  cross section.
The GDR width for ${}^{208}$Pb is approximately 2.5, 2.8, 3.0, 3.6 and 5.8 MeV for the case with constant nucleon-nucleon cross section of $\sigma_{NN}=0, ~50,~100, ~200~\text{mb}$ and the more realistic energy-dependent nucleon-nucleon cross section $\sigma_{NN}^\text{free}$, respectively.
These results confirm that GDR damping is sensitive to binary collisions, suggesting that the GDR width could be used to constrain the in-medium cross section.
Besides, the inset shows that the GDR width is still underestimated even with a constant nucleon-nucleon cross section of 200 mb, which is twice the upper limit of the {\it NN}  cross section in the original EQMD model.

\begin{figure}[htbp]
\resizebox{8.6cm}{!}{\includegraphics{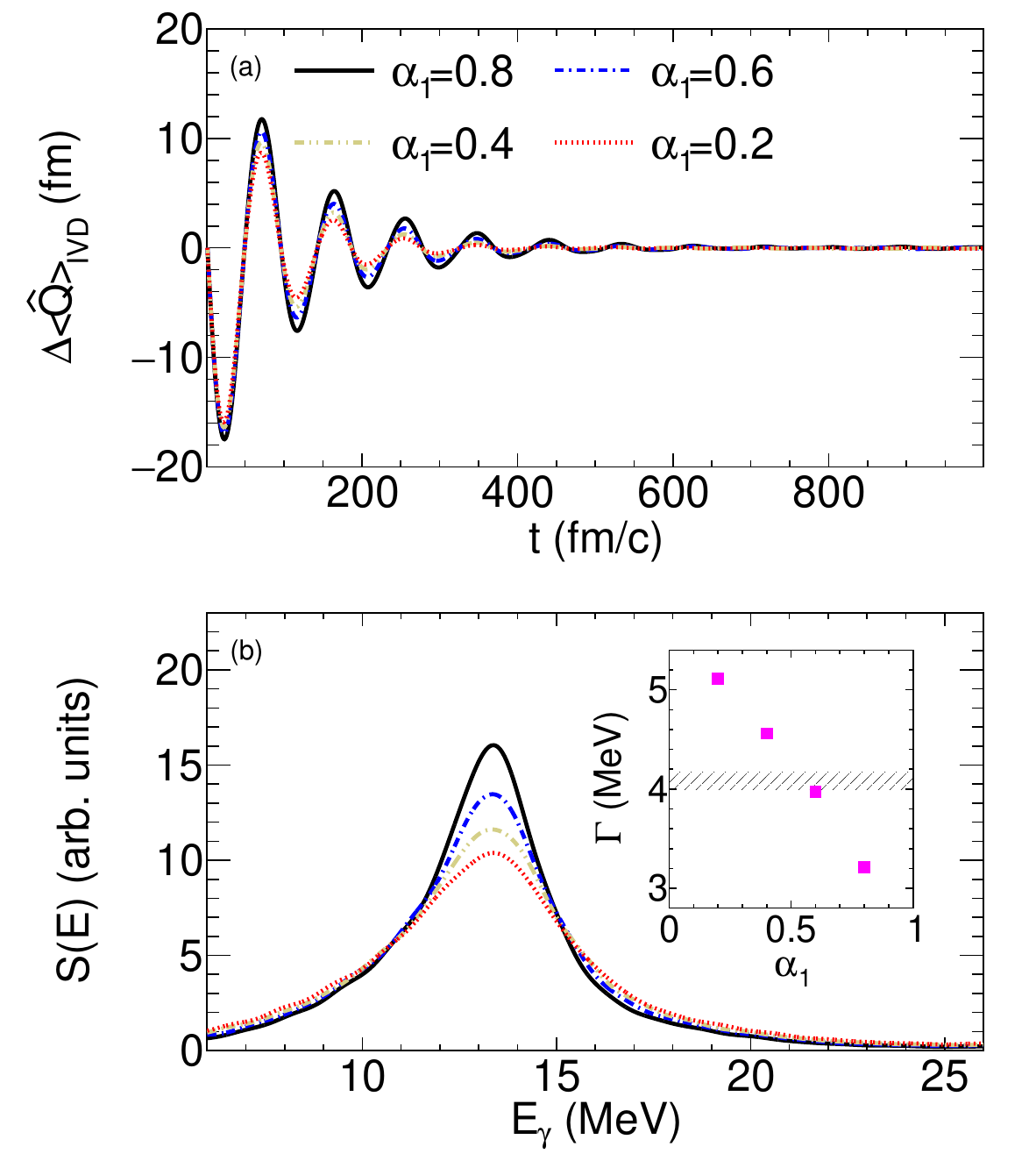}}
\caption{\label{fig:inmed} Same as Fig.~\ref{fig:different cross section}, but with different value of $\alpha_1$. }
\end{figure} 

\begin{figure}[htbp]
\resizebox{8.6cm}{!}{\includegraphics{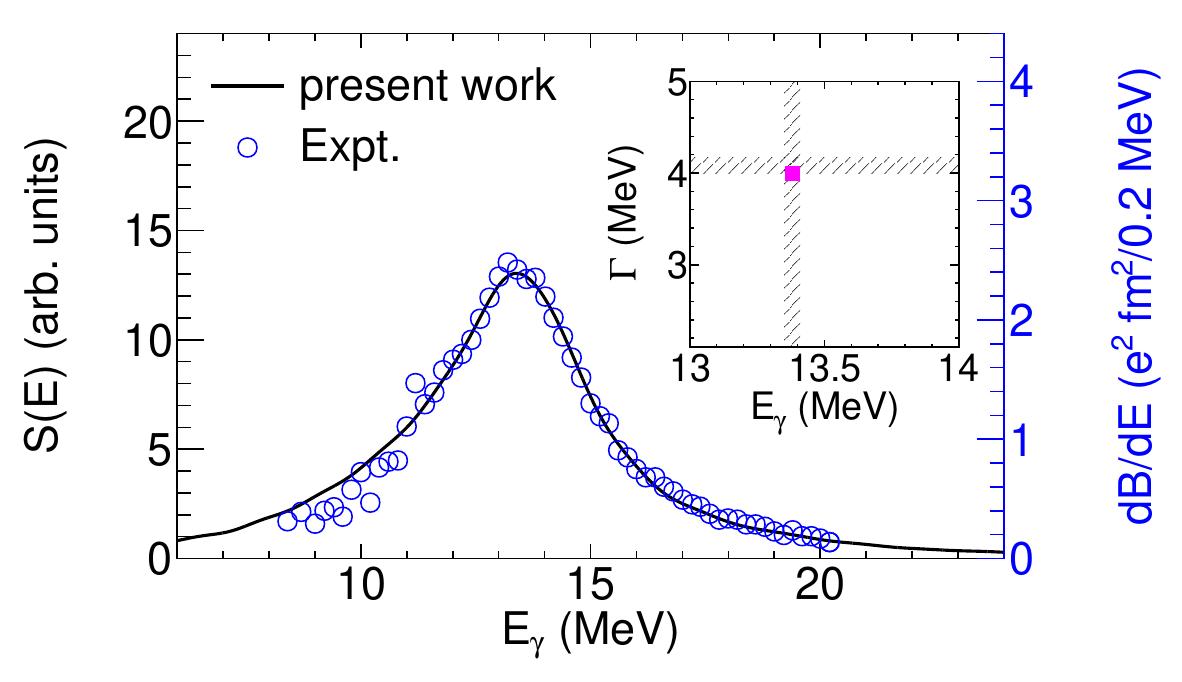}}
\caption{\label{fig:expt} 
Comparison of the GDR calculation for ${}^{208}$Pb (solid line, scaled by the left $Y$ axis) under the conditions $c_\text{s}=33.2~\text{MeV}$ and $\alpha_1$ = 0.57 against the experimental data (cycles, scaled by the right $Y$ axis) obtained from the RCNP~\cite{expt_data}.
The GDR peak and width is plotted as a square in the insert and the shaded area is the evaluated data.  
}
\end{figure}   

In the case of the free nucleon-nucleon cross section, the GDR width is significantly overestimated, as can be seen in Fig.~\ref{fig:different cross section}.
The primary reason for this lies in the lack of in-medium effects during binary collisions, when colliding nucleons are surrounded by a nuclear medium~\cite{wangrui2020}.
Here, we adopt the most common phenomenological parameterization of the cross section to introduce in-medium effects \cite{Klakow}, as follows:
\begin{equation}
\sigma_\text{NN}^\ast= \left(1-\alpha_1\frac{\rho}{\rho_0}\right) \sigma_\text{NN}^\text{free},
\end{equation}
where the parameter $\alpha_1$ reflects the strength of the medium correction.
$\rho_0$ is the nuclear saturation density and $\rho$ is the density at the collision vertex.
The upper panel (a) of Fig.~\ref{fig:inmed} shows the GDR oscillations obtained with different values of $\alpha_1$, and the lower panel (b) plots the corresponding strength functions.
As expected, the GDR damping decreases as the value of $\alpha_1$ increases.
At approximately $0.6$, the result is in good agreement with the evaluated GDR width.     

Fig.~\ref{fig:expt} displays the strength function of the GDR for ${}^{208}$Pb, including the collision term via the stochastic approach, as a solid line.
The cycles represent the experimental data from the reaction $^{208}\text{Pb}\left(\text{p}, \text{p}^{\prime} \right)$ in RCNP (see Ref.~\cite{expt_data}). 
In the inset, the peak position and width we calculated are plotted as a square, and the evaluated data obtained from Ref.~\cite{evaluation_data} are shown as a shaded area.
As can be seen from the figure, a value of  $c_\text{s}$  of about 33.2 MeV is required to reproduce the peak position of $E_\gamma = 13.38 \pm 0.03~\text{MeV}$.
Furthermore, a value of $\alpha_1$ as large as 0.57 is required to reproduce the GDR width of $\Gamma=4.08\pm0.09~\text{MeV}$, indicating a significant medium correction to {\it NN} scattering in the low-energy region.
Compared with the results of the original geometric approach shown in Fig.~\ref{fig:geometry}, the stochastic approach is clearly capable of providing a reasonable GDR decay width in $^{208}$Pb.
This suggests that the stochastic approach is a more appropriate method of treating the collision term when the {\it NN} cross section is very large.

\section{Conclusion\label{conclusion}}

In this article, we propose using a stochastic approach to model binary collisions, instead of the geometric approach employed in the original EQMD model.
Our results confirm that the peak position and width of the GDR are highly sensitive to the magnitude of the symmetry energy coefficient and the in-medium nucleon-nucleon cross section. This provides an opportunity to study the nuclear equation of state and the medium effect.
For the parameters $c_\text{s}=33.2~\text{MeV}$ and $\alpha_1=0.57$, the GDR calculation for  ${}^{208}$Pb is in good agreement with the evaluation data.
As with the LBUU model, our results also suggest a substantial decrease in {\it NN}  scattering in the low-energy region.

\section*{Acknowledgments}
This work is partially supported by the National Natural Science Foundation of China under Contracts No. $12405148$, $12347149$, $12547102$, and $12147101$, the National Key R\&D Program of China under Contract No. 2023YFA1606603, the Guangdong Major Project of Basic and Applied Basic Research No. 2020B0301030008, and the STCSM under Grant No. 23590780100 and 23JC1400200.

\end{CJK*}

\bibliography{stochastic_gdr}
\end{document}